\documentstyle[aps]{revtex}
\begin{document}
\draft
\title{Influence of resonant tunneling on the imaging of atomic defects 
on InAs(110) surfaces by low-temperature scanning tunneling microscopy}
\author{A. Depuydt and C. Van Haesendonck}
\address{Laboratorium voor Vaste-Stoffysica en Magnetisme, Katholieke Universiteit Leuven, B-3001 Leuven, Belgium}
\author{N.S. Maslova, V.I. Panov, V.V. Rakov and S.V. Savinov}
\address{Chair of Quantum Radio Physics, Moscow State University, 119899 Moscow, Russia}
\date{\today}
\maketitle
\begin{abstract}
We have used a low-temperature scanning tunneling microscope (STM) to study 
the surface of heavily doped semiconductor InAs crystals.  The crystals are 
cleaved in situ along the (110) plane.  Apart from atomically flat areas, we 
also observe two major types of atomic scale defects which can be identified
as S dopant atoms and as As vacancies, respectively.  The strong bias voltage 
dependence of the STM image of the impurities can be explained in terms of
resonant tunneling through localized states which are present near the
impurity.
\end{abstract}
\pacs{PACS numbers: 61.16.Ch, 68.35.Dv, 71.55.Eq, 73.20.Hb}
\begin{par}
With the advent of the scanning tunneling microscope (STM) and the decrease 
of system sizes down to the nanometer scale, experimental results can often 
no longer be interpreted in terms of the standard model for STM 
imaging~\cite{1}.  This is generally caused by the following reasons.  (i) 
For system sizes comparable to the atomic scale, the local density of states 
in the contact area can be strongly altered by the tunneling current 
(tip-sample interaction), resulting in the appearance of additional 
localized states near the Fermi level~\cite{2,3,4,5}.  (ii) Individual 
localized states start to dominate the tunneling current, because the radii 
of the localized states become comparable to the area of the contact 
size~\cite{6}.  (iii) In the presence of localized states, the finite 
relaxation rate of the nonequilibrium electrons has to be taken into 
account, especially at low temperatures where the relaxation rate may become 
smaller than the tunneling rate~\cite{7}.
\end{par}
\begin{par}
In a recent publication we could already show that the STM imaging and 
scanning tunneling spectroscopy of InAs(110) surfaces is strongly affected 
by tip induced band bending due to charges which are present on localized 
states at the tip apex~\cite{8}.  Here, we present a series of additional 
experimental results which demonstrate the role played by individual 
localized states related to atomic defects.  In order to identify this role, 
we have imaged dopant atoms and vacancies appearing at the surface of the 
InAs compound semiconductor at a temperature of $4.2 \, {\rm K}$.  All the 
experiments have been performed with a home built low-temperature STM with 
an in situ sample cleavage mechanism~\cite{9}.  The samples used in the 
experiments are n-type InAs monocrystals which have been heavily doped with 
S atoms ($n \simeq 5 \times 10^{17} \, {\rm cm}^{-3}$).  The crystals are 
cleaved in situ along the (110) plane.  Since the STM is cooled far below 
the boiling point of oxygen, the partial vapor pressure of oxygen is 
extremely low.  Therefore, surfaces such as GaAs(110) and InAs(110), which 
normally tend to oxidize very quickly, will stay clean under these 
conditions for many days.
\end{par}
\begin{par} 
Figure 1 shows a typical image of a InAs(110) surface.  The atomic rows can 
be clearly observed.  The images are similar to those obtained by Feenstra 
{\it et al.} on GaAs(110)~\cite{10}.  The image depends on the polarity of 
the applied bias voltage.  For negative sample bias voltage electrons 
tunneling out of the occupied surface states carry the current.  These 
states are located near the As lattice sites (Fig. 1a).  For positive sample 
bias voltage electrons tunnel into the empty states which are located near 
the In lattice sites (Fig. 1b).  For clarity the As sublattice in Fig. 1a 
has been copied on the In sublattice in Fig. 1b.
\end{par}
\begin{par}
Apart from the different energy dependence of the density of states for both 
type of atoms, relaxation of the surface atomic structure results in a tilt 
of the As atoms in the vertical direction.  The tilt is described in terms of 
a buckling angle and gives rise to a lateral shift between the two 
sublattices in the [001] direction which is easily observed in Fig. 1.  We 
also observe a shift in the [1\={1}0] direction which has been reported 
earlier for GaAs by Lengel {\it et al.}~\cite{11} and might be explained in 
terms of a slightly different tip-to-sample distance at different sample 
bias voltage.
\end{par}
\begin{par}
We have also obtained detailed STM images of single dopant atoms and 
vacancies at different polarities and values of the applied bias voltage and 
low constant  tunneling current ($\simeq 20 {\rm pA})$.  Figure 2 shows a 
series of STM images of an atomic impurity at different values of sample 
bias voltage.  From the observed frequency of this kind of atomic defects, 
we conclude that the defect shown in Fig. 2 is a single S dopant atom.  For 
negative bias voltage not exceeding a threshold value of about 
$-0.9 \, {\rm V}$, the STM image of the impurity appears as a depression 
(dark spot) between two neighboring atomic As rows (Fig. 2d).  When 
increasing the bias voltage above the threshold value, the impurity suddenly 
appears as a hillock (bright spot) rather than as a depression (Fig. 2a-c).  
When the bias voltage exceeds $-2.5 \, {\rm V}$, the bright spot 
corresponding to the impurity again disappears.  A comparable behavior is 
observed for positive bias voltage, with a different threshold value of 
about $+0.45 \, {\rm V}$ at which the impurity starts to appear as a 
hillock (Fig. 2e-h).
\end{par}
\begin{par}
In order to identify the origin of the pronounced bias voltage dependence of 
the STM image of the impurity in Fig. 2, we also imaged a neighboring 
defect at different applied bias voltage (Fig. 3).  From the appearance of 
this defect at different bias voltage, we conclude that it corresponds to a 
vacancy in the As sublattice~\cite{12}.  At more elevated bias voltage, the 
defect image does not suddenly change from a depression to a hillock.  On 
the other hand, at higher bias voltage the defect becomes surrounded by a 
cloud which indicates a decrease of the occupied density of states (Fig. 3f) 
and an increase of the empty density of states~\cite{13}.  In agreement with 
our experiments, we see that atomic defects of distinct nature behave 
different when imaged at different bias voltage.
\end{par}
\begin{par}
To understand qualitatively the changes in the STM image of a dopant atom 
we used a simple model which describes the effects in terms of switching 
on and off a channel for resonant tunneling (\cite{2} and see e.g.~\cite{14}).  Such 
channels are formed by localized states present near the impurities.  
Figure 4 illustratesthe tunneling process via a localized state.  The 
position $\varepsilon_{d}$ of the energy level associated with the localized 
state is assumed to depend on the applied bias voltage in a rather 
complicated way.  This dependence may be affected by an important 
modification of the unperturbed energy spectrum by the tip-sample 
interaction, charging effects~\cite{8} or the complex configuration of the 
electric field in the contact area.  For bias voltage smaller than the 
threshold value, a localized state does not participate in the resonant 
tunneling process, as illustrated in Fig. 4a (initial position 
$\varepsilon_{d}^0$).  The impurity appears as a dark spot in the STM image, 
indicating an important reduction of the local density of states near the 
impurity.
\end{par}
\begin{par}
The threshold value $V_{th}$ of the bias voltage, above which resonant 
tunneling sets in, can be estimated by relying on the simple condition
\begin{equation}
{\varepsilon_{d}={{\varepsilon_{d}^0}+{f(V_{th})}={E_{F}}}}\label{one}
\end{equation}
where $\varepsilon_{d}^0$ corresponds to the unperturbed position of the 
localized impurity level.  The function $f(V)$ describes the dependence of 
the position of the impurity level on applied bias voltage.  A resonant 
tunneling current will start to flow for bias voltage larger or equal than 
$V_{th}$, i.e., when $\varepsilon_{d}$ is located between $E_{F}$ and 
$E_{F}-eV$ (Fig. 4a, position 2 of the level $\varepsilon_{d}$).  When the 
impurity level $\varepsilon_{d}$ is located above the conduction band edge, 
the current is determined by the tunneling rate.  Because of the heavy 
n-type doping, free electron states are available between the conduction 
band edge $E_{c}$ and the Fermi level $E_{F}$.  We assume that the 
relaxation rate for the nonequilibrium electrons on the localized state 
exceeds the tunneling rate for the energy levels in the conduction band.  
When further increasing the applied bias voltage, $\varepsilon_{d}$ is 
located within the semiconductor band gap (Fig. 4a, position 3 of the level 
$\varepsilon_{d}$) and the resonant tunneling current will now be determined 
by the relaxation rate for the electrons occupying the localized level.  
This relaxation rate is smaller than the tunneling rate for energies within 
the band gap~\cite{15}.  Therefore, the tunneling current decreases and the 
bright spot corresponding to the impurity image disappears (Fig. 2).
\end{par}
\begin{par}
As already indicated above, the exact location of the energy level 
$\varepsilon_{d}$ relative to the band gap edges is strongly affected by 
charging effects, tip-induced band bending and modifications of the initial 
density of states by the tunneling processes.  Therefore, a complicated 
and possibly nonmonotonic dependence of the location of the energy level 
$\varepsilon_{d}$ on the applied bias voltage $V$ will emerge, resulting in 
a switching on and off of the resonant tunneling channel.  Moreover, it is 
possible that there is more than one localized state connected to the 
impurity.  For positive applied bias voltage the situation remains very 
similar (Fig. 4b).  For a complicated dependence $f(V)$ of the energy level 
$\varepsilon_{d}$, one single impurity level can be responsible for switching 
on and off the resonance channel at different threshold values $V_{th}$ for 
positive and negative bias voltage.  For positive bias voltage the resonant 
tunneling current starts to flow when $\varepsilon_{d}$ is located between 
$E_{F}$ and $E_{F}-{eV}$.
\end{par}
\begin{par}
Other possible mechanisms for switching on or off the resonant tunneling 
channel may be related to changes of the effective potential well near the 
impurity which are induced by the applied bias voltage.  In quantum mechanics 
the presence of localized states in a symmetric quantum well is well 
known~\cite{16} and there exists a critical value for the asymmetry at which 
the localized states disappear.  The applied bias voltage changes the 
symmetry of the effective potential well.  Therefore, the localized energy 
levels present in the well can disappear at high bias voltage.
\end{par}
\begin{par}
In conclusion, we have found that the STM images of an individual atomic 
defect (doping atoms) on an InAs(110) surface strongly depend on the value of 
the applied tunneling bias voltage.  Moreover, there exist threshold values 
for both positive and negative polarities of the tunneling bias voltage.  
When the bias voltage reaches these threshold values, the STM image of the 
defect drastically changes from appearing as a depression (low density of 
states) towards a hillock (high density of states).  We can explain the 
observed behavior in terms of the formation of resonant tunneling channels 
which are connected to localized energy states near the atomic defect.  The 
exact position of these localized states is determined by tip-sample 
interactions, charging effects and relaxation processes which all depend on 
the applied bias voltage.  The possibility to have resonant tunneling may 
therefore depend in a very complicated and even nonmonotonic way on the 
applied bias voltage.  Our model supports the idea that in low-temperature 
STM experiments, the nanometer size of the tunneling contact strongly 
enhances the influence of individual localized states.
\end{par}                                      
\acknowledgments
The work at Leuven has been supported by the Fund for Scientific Research - 
Flanders (FWO) as well as by the Flemish Concerted Action (GOA) and the 
Belgian Inter-Universitary Attraction Poles (IUAP) research programs.  The 
work in Moscow has been supported by the Russian Ministry of Research 
(Surface Atomic Structures, grant 95-1.22; Nanostructures, grant 1-032) and 
the Russian Foundation of Basic Research (RFBR, grants 96-0219640a and 
96-15-96420).  The collaboration between Moscow and Leuven has been funded by 
the European Community (INTAS, project 94-3562).

\begin{par}
{\bf Fig. 1.} Constant-current STM images acquired at sample bias voltage of (a)
$-1.5 \, {\rm V}$ and (b) $+1.5 \, {\rm V}$. The scanned area is $26 \, 
{\rm \AA} \times 14 \, {\rm \AA}$ and the tunnel current is fixed at $20 \, 
{\rm pA}$.
\end{par}
\begin{par}
{\bf Fig. 2.} Constant-current STM images of a dopant atom on the InAs(110) 
surface acquired at different sample bias voltage. The scanned area is $44 \, 
{\rm \AA} \times 44 \, {\rm \AA}$ and the tunnel current is fixed at $20 \, 
{\rm pA}$.
\end{par}
\begin{par}
{\bf Fig. 3.} Constant-current STM images of a As vacancy on the InAs(110) 
surface acquired at different sample bias voltage. The scanned area is $70 \, 
{\rm \AA} \times 70 \, {\rm \AA}$ and the tunnel current is fixed at $20 \, 
{\rm pA}$.
\end{par}
\begin{par}
{\bf Fig. 4.} Schematic diagram for localized state assisted tunneling (a) 
for negative sample bias voltage, and (b) for positive sample bias voltage.
\end{par}
\end{document}